\documentclass[3p,times,twocolumn,number,longtitle]{elsarticle}

\usepackage{graphicx}

\usepackage[version=4]{mhchem}
\usepackage[mathlines]{lineno}%

\usepackage{verbatim}
\usepackage{xspace}
\usepackage{multirow}
\usepackage[normalem]{ulem}
\usepackage[colorlinks=true]{hyperref}
\usepackage[usenames, dvipsnames, table]{xcolor} 
\usepackage{placeins}

\newcommand{\0}{\ensuremath{0\nu\beta\beta}}
\newcommand{\2}{\ensuremath{2\nu\beta\beta}}

\newcommand{\Q}{\ensuremath{Q_{\beta\beta}}}

\biboptions{sort&compress}

\begin{document}

\begin{frontmatter}

\title{Event Reconstruction in a Liquid Xenon Time Projection Chamber with an Optically-Open Field Cage}

\author[LLNL]{T.~Stiegler\fnref{cor}}
\fntext[cor]{Corresponding author \textit{stiegler1@llnl.gov}}
\author[LLNL]{S.~Sangiorgio}
\author[LLNL]{J.P.~Brodsky}
\author[LLNL]{M.~Heffner}
\author[McGill]{S.~Al Kharusi}
\author[Erlangen]{G.~Anton}
\author[PNNL]{I.J.~Arnquist}
\author[Carleton]{I.~Badhrees\fnref{KingAbdulaziz}}
\author[Duke]{P.S.~Barbeau}
\author[UIUC]{D.~Beck}
\author[ITEP]{V.~Belov}
% ...
\address[LLNL]{Lawrence Livermore National Laboratory, Livermore, CA 94550, USA}

\address[McGill]{Physics Department, McGill University, Montr\'eal, Qu\'ebec H3A 2T8, Canada}
\address[Erlangen]{Erlangen Centre for Astroparticle Physics (ECAP), Friedrich-Alexander University Erlangen-N\"urnberg, Erlangen 91058, Germany}
\address[PNNL]{Pacific Northwest National Laboratory, Richland, WA 99352, USA}
\address[Carleton]{Department of Physics, Carleton University, Ottawa, Ontario K1S 5B6, Canada}
\address[Duke]{Department of Physics, Duke University, and Triangle Universities Nuclear Laboratory (TUNL), Durham, NC 27708, USA}
\address[UIUC]{Physics Department, University of Illinois, Urbana-Champaign, IL 61801, USA}
\address[ITEP]{Institute for Theoretical and Experimental Physics named by A. I. Alikhanov of National Research Center ``Kurchatov Institute'', Moscow 117218, Russia}
% ...

\fntext[KingAbdulaziz]{also at King Abdulaziz City for Science and Technology, Riyadh, Saudi Arabia}
\fntext[alsoSNOLAB]{also at SNOLAB, Ontario, Canada}
\fntext[alsoCAS]{also at University of Chinese Academy of Sciences, Beijing, China}
\fntext[nowIntel]{now at Intel, Portland, OR, USA}
\fntext[nowYale]{now at Yale University, New Haven, CT, USA} 

\begin{abstract}
nEXO is a proposed tonne-scale neutrinoless double beta decay (\0) experiment using liquid \ce{^136Xe} (LXe) in a Time Projection Chamber (TPC) to read out ionization and scintillation signals. Between the field cage and the LXe vessel, a layer of LXe (``skin'' LXe) is present, where no ionization signal is collected. Only scintillation photons are detected, owing to the lack of optical barrier around the field cage. In this work, we show that the light originating in the skin LXe region can be used to improve background discrimination by 5\% over previous published estimates. This improvement comes from two elements. First, a fraction of the $\gamma$-ray background is removed by identifying light from interactions with an energy deposition in the skin LXe. Second, background from \ce{^222Rn} dissolved in the skin LXe can be efficiently rejected by tagging the $\alpha$~decay in the \ce{^214Bi}-\ce{^214Po } chain in the skin LXe. %

\end{abstract}


\end{frontmatter}

\section{\label{sec:intro}Introduction}
%\linenumbers
    Neutrinoless double beta decay (\0) is a second-order weak transition that is predicted to occur in several even-even nuclei~\cite{PDG2016} if neutrinos are Majorana fermions. The observation of this process would indicate lepton number violation, demonstrate the Majorana nature of neutrinos~\cite{Majorana:1937}, and provide valuable information about the absolute scale of the neutrino mass spectrum. 
    
    nEXO is a proposed tonne-scale detector that will use $\sim$5000 kg of isotopically enriched liquid xenon (LXe) in a cylindrical Time Projection Chamber (TPC) to search for \0 in \ce{^136Xe}. A detailed sensitivity analysis~\cite{Albert:2017hjq} has been published along with a pre-conceptual design of nEXO \cite{Kharusi:2018eqi}. 
    nEXO measures ionization and scintillation  signals~\cite{exo-200_anti_correlation} from the LXe volume inside a field cage of evenly-spaced field-shaping rings that establish the required TPC drift electric field. The scintillation light is collected by silicon photomultipliers (SiPMs) on the cylindrical ``barrel” of the detector and the ionization electrons by charge collection tiles on the anode. The field-cage assembly is surrounded by an insulating layer of LXe, referred to as the “skin LXe”, where no ionization signal can be collected. A scintillation signal, on the other hand, is collected from interactions both inside and outside the field cage. In nEXO, approximately 25\% of the liquid xenon is located in the skin LXe region. The question addressed in this paper is to identify the best way to handle the scintillation light collected from the skin LXe, and to demonstrate that it can be used to improve background discrimination.
    
\begin{figure}[tbp]
        \centering
        \includegraphics[width=\columnwidth]{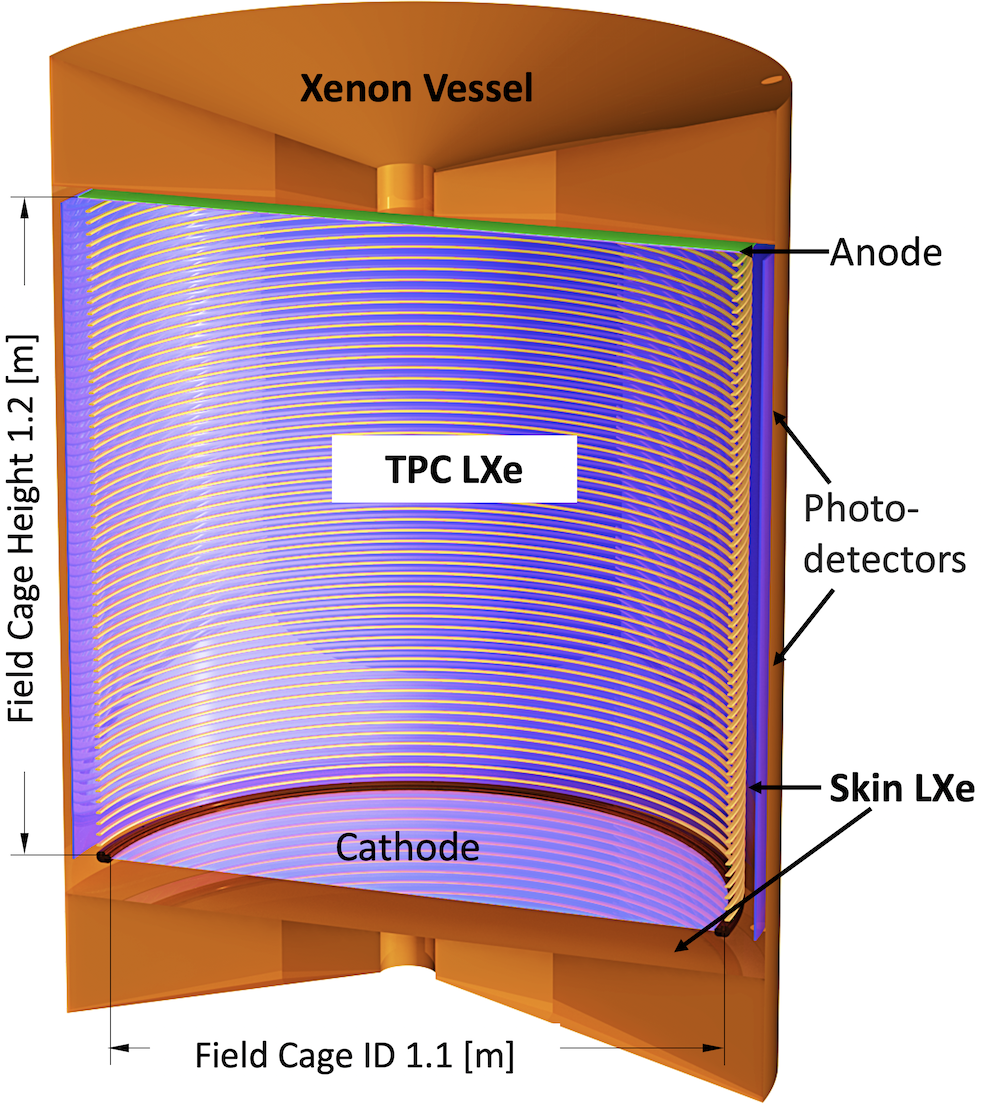}
        \caption{Cross-section of the proposed TPC for nEXO.}
        \label{fig:nexo_tpc_cross}
    \end{figure}

    The field cage surrounded by an insulating skin of LXe is a typical design feature in detectors of this type, but different approaches have been taken with regard to light collection in the skin LXe. One approach has photodetectors installed on the ends of the cylindrical volume and a way to optically isolate the LXe volume inside the field cage from the skin LXe outside~\cite{Aprile:2017aty, Mount:2017qzi, exo-200_detector_paper}. We refer to this arrangement, with two distinct optical volumes, as a “closed-field-cage” design. %
    nEXO dispenses with the optical barrier around the TPC field cage, and is therefore an "open-field-cage" design with a single optical volume. Fig.~\ref{fig:nexo_tpc_cross} shows a cross-section of the nEXO detector.
    
    By dispensing with the optical shield, nEXO removes a potential source of radioactive and chemical contamination and improves LXe circulation. In this paper, we show that the absence of a light barrier around the field cage does not prevent the use of light originating in the skin LXe region to classify events. It can be used to improve background discrimination by 5\% over previous estimates~\cite{Albert:2017hjq}. 
    
    The previous estimates did not evaluate the implications of the open-cage configuration but rather made two assumptions. First, it was assumed that $\alpha$ particles in the skin LXe, which produce intense scintillation signals, could often be detected and identified. A rejection efficiency of 40\% was used for \ce{^214Bi} decays in the skin LXe and on surfaces using the time-correlated \ce{^214Po} $\alpha$ (the ``Bi-Po'') decay pair (half-life of 164~$ \mu$s~\cite{Basunia:2014}). Second, it was assumed that $\gamma$-ray and $\beta$ interactions in the skin LXe would neither produce an exploitable signal, nor confuse coincident signals from interactions inside the TPC. \par
    This work reevaluates the above assumptions using the open-field-cage nature of nEXO TPC for event reconstruction and background discrimination. First, the analysis quantifies the tagging efficiency for Bi-Po decays in the skin LXe. Second, simultaneous scintillation light in the skin LXe and inside the TPC is modeled in detail, showing that this light can be exploited to reject coincident skin $\gamma$-ray interactions based on their multi-sited nature. In both cases, the resulting new background estimates for nEXO are compared to those obtained with the assumptions from \cite{Albert:2017hjq}.

\section{\label{sec:sims}Methodology}
    
    \subsection{\label{sec:geo}Detector simulations}
        The impact of the open-field-cage design was studied using a \textsc{Geant4}-based Monte Carlo simulation ~\cite{GEANT4-2003} implemented according to the nEXO detector geometry described in \cite{Kharusi:2018eqi} and summarized below. Particle transport and energy deposition were determined using the same parameters used in~\cite{Albert:2017hjq}. 
        
\begin{figure}[tbp]
        \centering
        \includegraphics[width=\columnwidth]{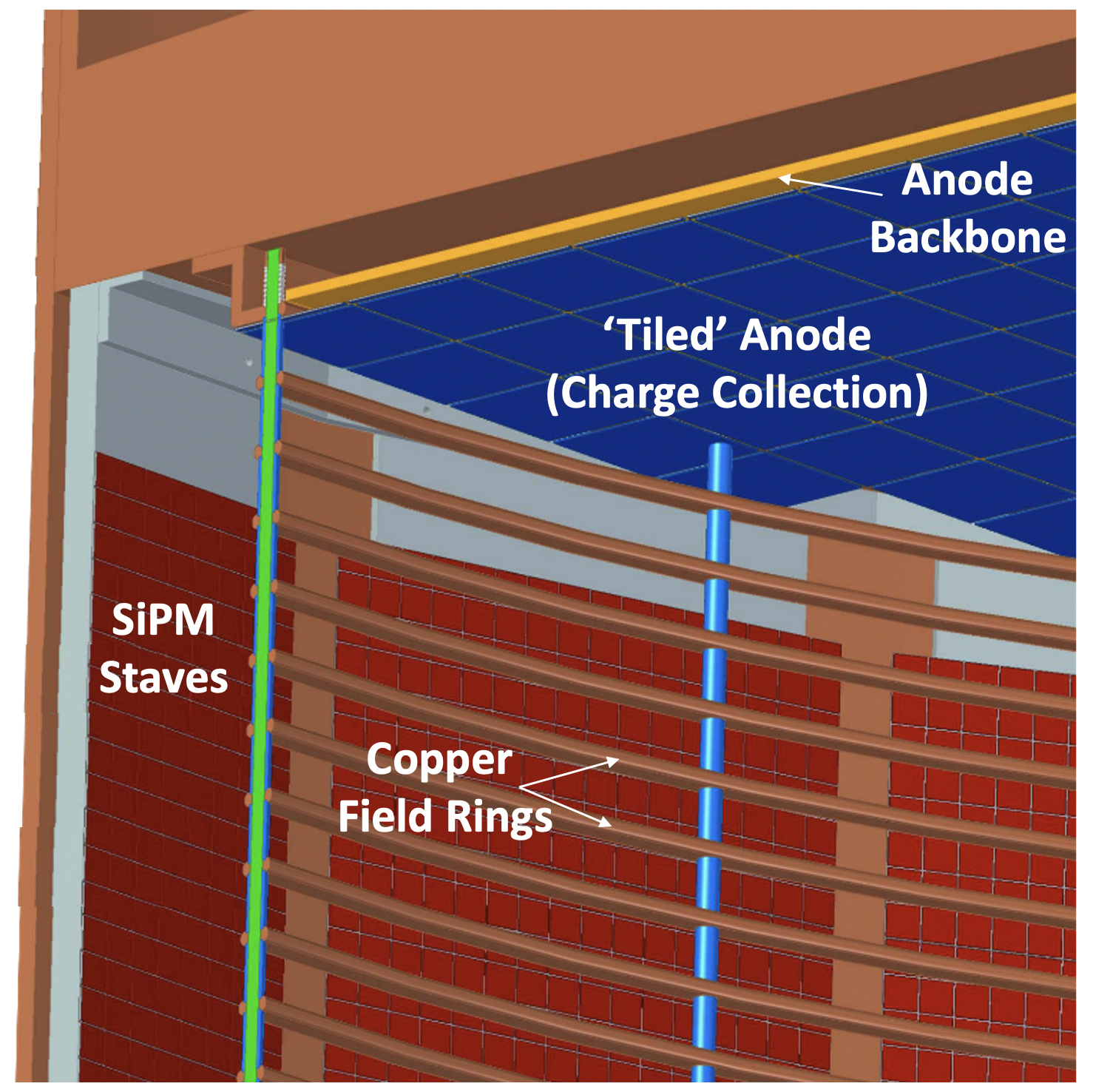}
        \caption{Cutaway sketch of the anode section of the TPC. The configuration of the charge collection tiles is shown below a cutaway of the anode backbone.~\cite{Kharusi:2018eqi}}
        \label{fig:anode_closeup}
        \end{figure}

        The simulated TPC consists of 4896~kg of homogeneous LXe, isotopically enriched to 90\% in the \ce{^136}Xe nuclide. %
        The field gradient is controlled by the voltage difference between the anode and the cathode, and a system of field-shaping rings assures parallel field lines to enhance charge collection. These components are shown in Fig.~\ref{fig:anode_closeup}. 
        The TPC bottom is an opaque cathode at high potential, and the top is an anode at ground potential. In nEXO, about 25\% of the LXe is located outside the field cage ($~5\%$ under the cathode, $~1\%$ above the anode). The small fraction of optically-inactive xenon (0.6\%) behind the SiPM staves was neglected in this work.
        
        Charge is collected by an array of fused silica ``tiles" covered electrodes. Fig.~\ref{fig:anode_closeup} shows the placement of the tiles on the underside of the anode surface. All ionization produced by interactions in the LXe inside the field rings is collected on the tiles, no ionization is collected from outside the field rings. 
        
        The scintillation light is collected by an array of SiPMs sensitive to the 175~nm wavelength xenon scintillation light~\cite{Jamil:2018tkx,Gallina2019}. The SiPMs are arranged in 24 staves outside the field rings, as shown in Fig.~\ref{fig:anode_closeup}, for a total photosensitive surface area of $\sim4$~m$^2$. 
        The top of the staves are inline with the anode, and the bottoms extend 6~cm below the cathode. The SiPMs collect scintillation light from particle interactions both inside and outside of the TPC field cage which has an optical transparency of 79\%. 
        The nontransparent cathode and anode reduce direct line of sight from the photosensors to light produced above the anode or below the cathode.

    \subsection{\label{sec:sim} Event generation}
        \ce{^222Rn}, and its radioactive daughters, are well-known liquid xenon contaminants that have been studied in detail~\cite{Albert:2015nta, Albert:2015vma}. Of particular concern for nEXO is the \ce{^222Rn}-daughter, \ce{^214Bi}, whose decay includes a $\gamma$~ray with an energy of $2447.7$~keV and a branching ratio of 1.5\% \cite{Basunia:2014}, that is only 10~keV away from the \ce{^136Xe} \0 Q-value (\Q$=2458.07\pm0.31$ keV) \cite{PhysRevLett.98.053003,McCowan:2010zz}. The \ce{^214Bi} $\beta$ decay is followed by a \ce{^214Po} $\alpha$~decay with a half-life of 164~$ \mu$s~\cite{Basunia:2014}. By tagging the $\alpha$~decay, this time-correlated Bi-Po decay chain can be identified, and the resulting background events rejected. 
        
        Bi-Po decay chains were simulated both as diffused sources in the LXe volume as well as localized sources on certain surfaces. The simulation was performed in this way because both neutral and charged daughters are present in the \ce{^222Rn} decay chain~\cite{Albert:2015vma}. The neutral daughters remain in the bulk LXe volume while the charged drift in the electric field and plate onto negatively charged surfaces. 
        The ions in the barrel of the skin LXe plate onto the outer radial surface of the field rings, and those inside the TPC field cage plate on the top surface of the cathode. The charged daughters in the skin LXe above the anode are assumed to remain in the bulk LXe because the electric field in that region is zero. Under the cathode, any ions plate onto the underside surface of the cathode.
        The number of \ce{^214Bi} decays from each region was calculated as in \cite{Albert:2015vma} using measured ion fractions, mobility, and drift time with the expected electric fields in LXe for nEXO. This resulted in 15.6\% of the \ce{^214Bi} decays taking place in the TPC LXe, 5.4\% in the skin LXe, 58.9\% on the cathode surface, and 20.1\% on the field rings outer surface. A steady state population of 600 \ce{^222Rn} atoms in the total LXe volume of nEXO was assumed~\cite{Kharusi:2018eqi}. %
        
        A uniform spatial distribution was assumed for \ce{^214Bi} decays from \ce{^222Rn} daughters that remain in the liquid xenon. Ionized \ce{^214Bi} daughters were simulated on the nearest corresponding surface, at zero depth in the material. The subsequent \ce{^214Po} nuclei were allowed to remain on the surface or move into the LXe according to \textsc{Geant4}'s physics model for ion transport. 
        
        To study the impact of the open-field-cage design on coincident multi-site $\gamma$-induced backgrounds, interactions from the decays of \ce{^238U} and \ce{^232Th} and their daughters were simulated. These radionuclides constitute the dominant background in nEXO. They are present as contamination in detector materials, with the largest contribution from the copper that makes up the TPC vessel. %
        As in \cite{Albert:2017hjq}, the radionuclides simulated for the \ce{^238U} and \ce{^232Th} decay chains were selected based on the emitted $\gamma$-radiation energy $>100$~keV and intensity $> 1\%$. The resulting energy deposits in the skin LXe and TPC LXe are combined with appropriate branching ratios. The event rates assume \ce{^238U} and \ce{^232Th} decay chain equilibrium. 
        
        Using \textsc{Geant4}, each of the selected decay types were simulated in sufficient quantity to obtain statistically significant numbers of events in the inner 2000~kg LXe volume. The decay type, location, and quantity simulated are listed in Table~\ref{tab:sims}. The same Monte Carlo events are used for both the open- and closed-field-cage analyses.
        \begin{table}[tbp]
            \centering
            \begin{tabular}{llc}
            \hline
            \hline
            \textbf{Event Type} & \textbf{Location} &\textbf{\# Decays}\\ 
            \hline
                \ce{^238U} Chain & TPC Vessel & $10^8$  \\
                \ce{^238U} Chain & Internal TPC Components & $10^7$  \\
                \ce{^232Th} Chain & TPC Vessel & $10^8$  \\
                \ce{^232Th} Chain & Internal TPC Components & $10^7$  \\
                \ce{^214Bi} -\ce{^214Po} & Skin LXe Volume & $10^8$ \\
                \ce{^214Bi} -\ce{^214Po} & Outer Field Ring Surface & $10^8$  \\
            \hline    
            \hline
            \end{tabular}
            \caption{Event type, location, and number of simulated primary decays for this study.}
            \label{tab:sims}
        \end{table}

    \subsection{\label{sec:event} Event reconstruction in an open-cage TPC design}
        The output of the \textsc{Geant4} simulation is processed to apply detector response effects. Simulated events are analyzed  to extract the event parameters of interest:  multiplicity (Single-Site (SS) or Multi-Site (MS)), distance from the nearest TPC surface (Standoff distance), and reconstructed event energy. This analysis largely follows the procedure detailed in Ref.~\cite{Albert:2017hjq}, although some key differences will be called out in the following summary.
    
        The multiplicity of the simulated events is determined first. An algorithm groups energy deposits within 3~mm of each other into clusters. The 3~mm cluster size is chosen to emulate the discrimination ability projected for nEXO using a separate detailed charge transport simulation ~\cite{Li_2019}. To be labeled SS, an event can only have one reconstructed cluster inside the TPC LXe. This requirement separates the \2 and \0 events (which are primarily SS) from other types of interactions, like Compton scatters, which are typically MS. Any event that has more than one reconstructed cluster is labeled as MS. For this study, SS events with one or more additional energy deposits in the skin LXe are further labeled as MS-skin events. This MC-truth value is stored to later assess the efficiency of the analysis. 
   
        Once all the energy deposits are identified and recorded the production of scintillation and ionization quanta for each are explicitly calculated. NEST version 2.1~\cite{szydagis_m_2018_1314669,jason_brodsky_2019_2535713} is used to compute the correlated light and charge quanta based on the electric field at a given location, and the deposited energy. The electric field is based on a simplified version of the detailed COMSOL-based ~\cite{comsol:2019} electrostatic model of nEXO.
        
        \begin{figure}[tbp]
            \centering
            \includegraphics[width=0.99\columnwidth]{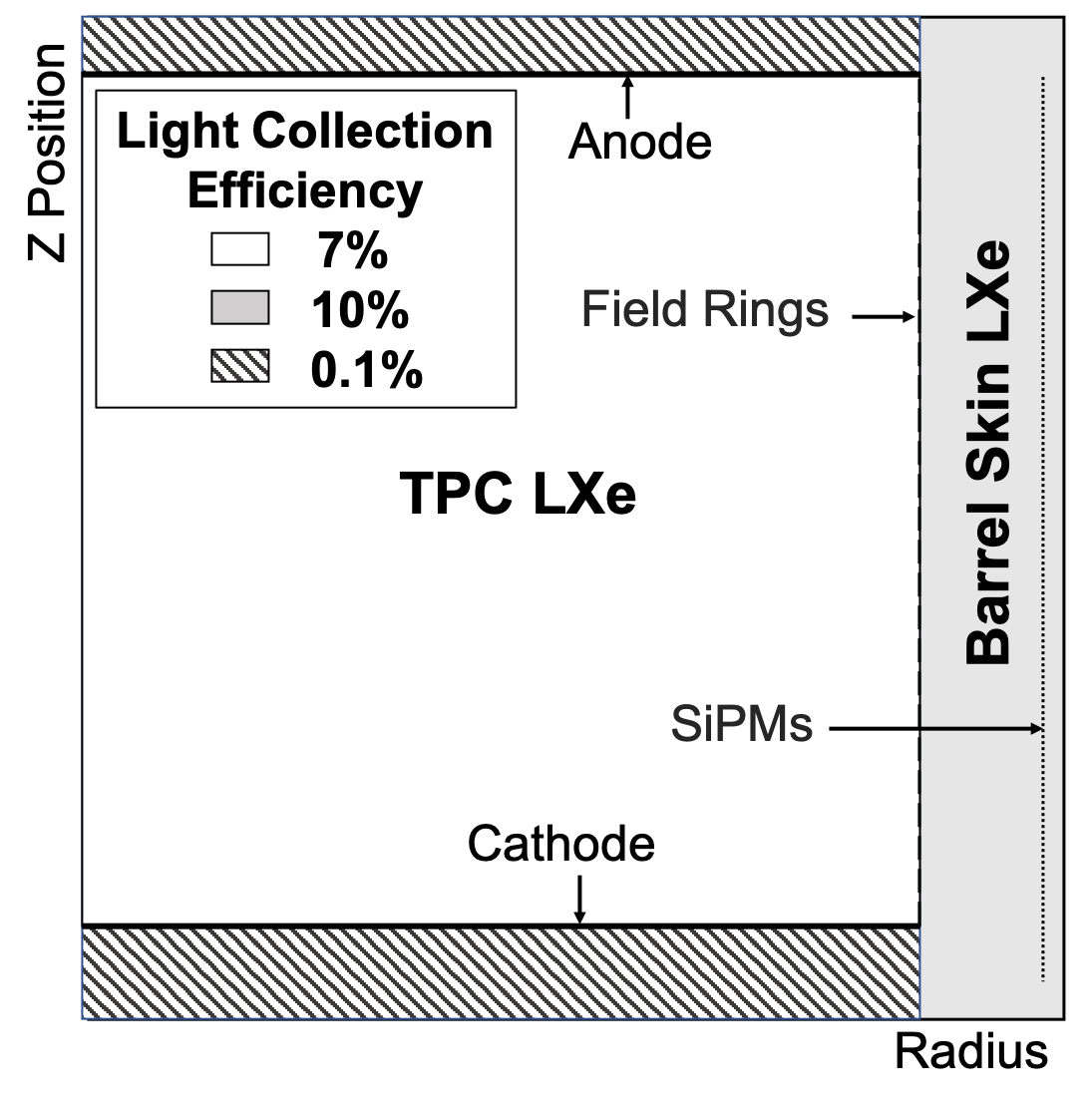}
            \caption{Schematic map of the estimated light collection efficiency applied to simulations as a function of position (Z vs Radius). These estimated efficiencies are uniform across each region, an approximation based on \cite{Albert:2017hjq,Kharusi:2018eqi}. The TPC LXe is set at 7\%, the barrel of the skin LXe at 10\%, and the regions above the anode and below the cathode at 0.1\%.}
            \label{fig:lceMap}
        \end{figure}
        
        The charge signal is lost in the skin LXe and the light collection efficiency varies depending on position, primarily for geometrical reasons. %
        The estimated light collection efficiencies are uniform across each region, an approximation based on previous work \cite{Albert:2017hjq,Kharusi:2018eqi} and verified through dedicated \textsc{Geant4} optical transport simulations of the skin LXe. A coating of Al$ + $MgF$_{2}$ boosts the reflectivity of the cathode and field shaping rings to 80\%, increasing the overall light collection efficiency. The TPC LXe is set at 7\%, the barrel of the skin LXe at 10\%, and the regions above the anode and below the cathode at 0.1\%. A schematic representation is given in Fig~\ref{fig:lceMap}. Charge and light collection efficiencies are applied to the number of scintillation and ionization quanta from NEST. %
    
        After all the energy deposits in an event are divided into collected light and charge, the event energy is calculated following the technique in~\cite{Albert:2013gpz} where the anti-correlation between light and charge signals in LXe~\cite{exo-200_anti_correlation} is leveraged to generate a rotated energy axis with significantly better resolution than the individual channels. The effects of instrumental noise in the light and charge signals are added during this analysis, with noise values chosen to match the nEXO expected energy resolution at \Q~of $1\%$~\cite{Kharusi:2018eqi}. 
        
        In Ref.~\cite{Albert:2017hjq}, the full description of the complex experimental backgrounds was distilled down into a single reference value -- namely the number of background events reconstructed within the FWHM window around \Q~in the inner 2000~kg LXe volume. The same simplifying metric is used in this analysis.
        
        \begin{figure}[tbp]
            \centering
            \includegraphics[width=1.0\columnwidth]{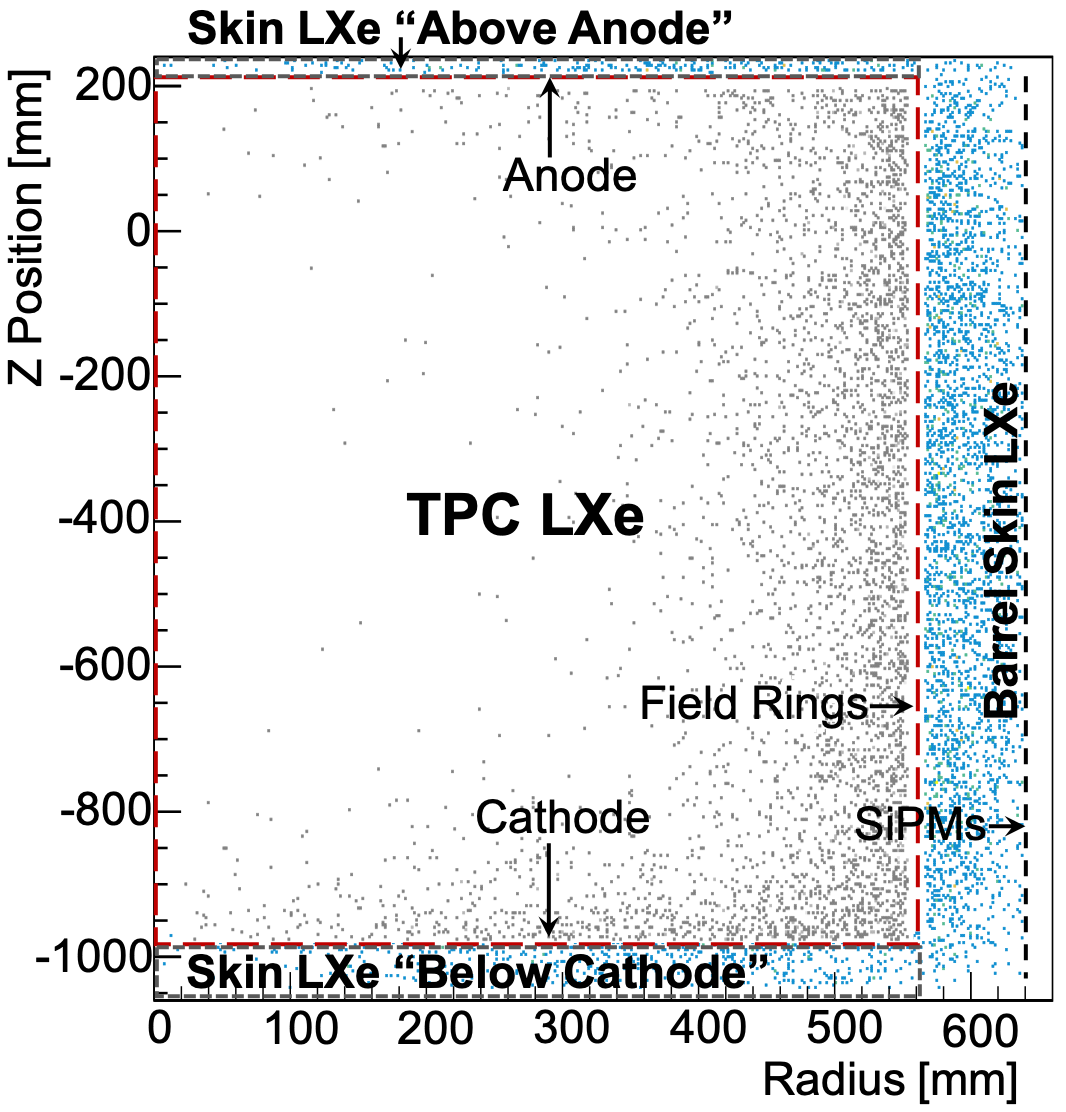}
            \caption{Position of time correlated Bi-Po interactions (Z vs Radius) produced in the skin LXe. Reconstructed SS events in the TPC LXe from \ce{^214Bi} coincident $\gamma$~rays are shown in the TPC LXe (grey). The position of coincident $\alpha$~decays from \ce{^214Po} in the skin LXe are shown in the skin LXe (blue). The SS event in the TPC LXe is background if the time correlated $\alpha$~decay cannot be identified.}
            \label{fig:timeBiPo}
        \end{figure}

    \subsection{\label{sec:biposkin}Analysis of \texorpdfstring{\ce{^214B\lowercase{i}} -\ce{^214P\lowercase{o}}}{Bi-Po} decays in the skin LXe}
        There are four regions that contribute \ce{^214Bi}-\ce{^214Po} backgrounds: the TPC LXe, the skin LXe, the outer surface of the field rings, and the top surface of the cathode. Backgrounds from the \ce{^214Bi} decays can be vetoed and rejected if the \ce{^214Po} $\alpha$~decay is detectable. Fig.~\ref{fig:timeBiPo} shows the positions of both the Bi-Po decays and SS energy deposits inside the TPC LXe produced by $\gamma$~rays from  \ce{^214Bi} $\beta$~decays in the skin LXe.
        
        Interactions of $\alpha$~particles inside the TPC LXe are easily distinguished from $\beta$ and $\gamma$-ray interactions based on the ratio of charge and light signals which differ by more than a factor of 20~\cite{Albert:2015vma}. 
        
\begin{figure}[tbp]
            \centering
            \includegraphics[width=0.99\columnwidth]{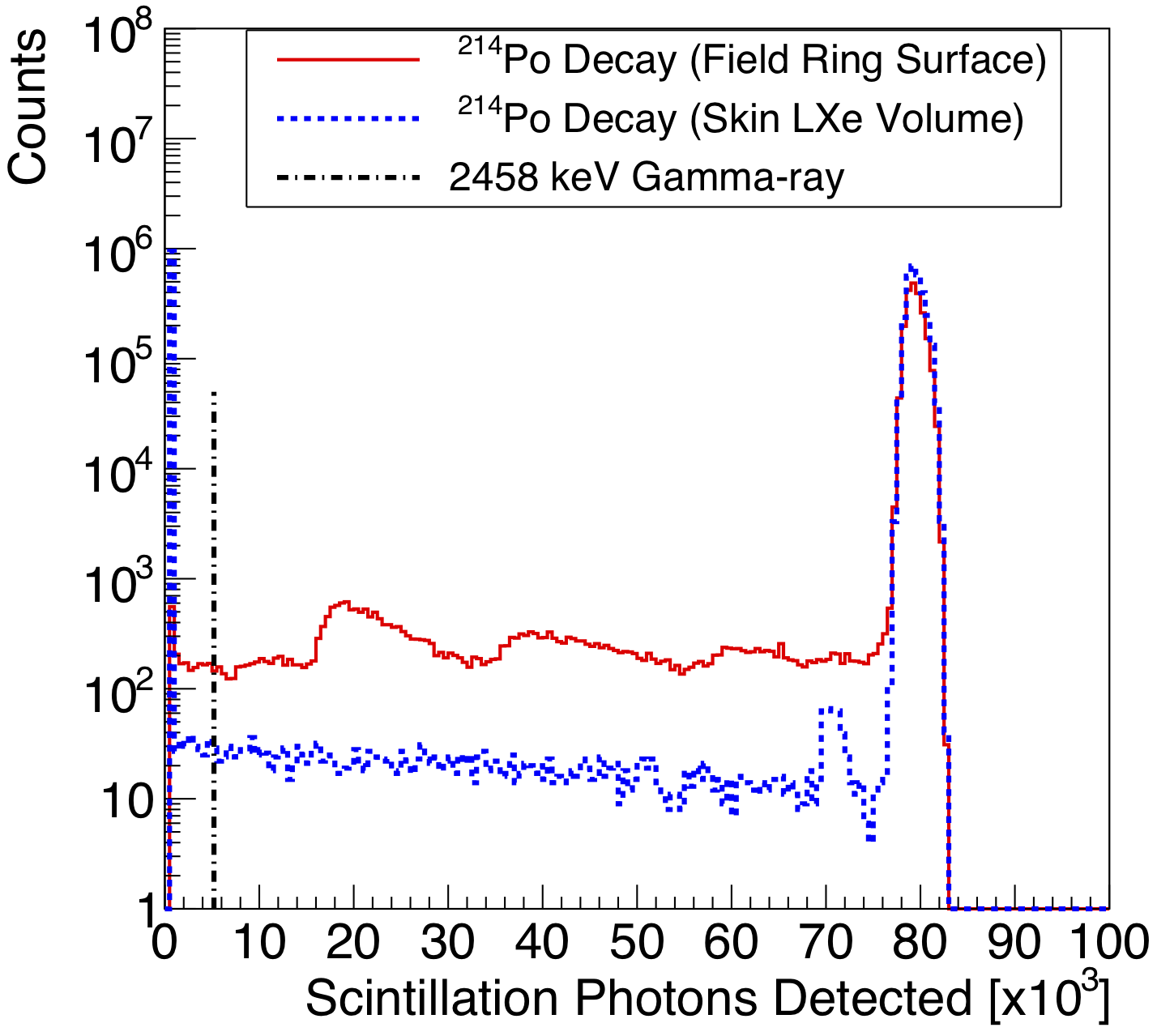}
            \caption{Number of photons detected from \ce{^214Po} $\alpha$~decays in the skin LXe (dotted) and on the field ring surfaces (solid). The dot-dashed line marks the maximum light detected from a $\gamma$~ray with energy at \Q~in the barrel of the skin LXe.}
            \label{fig:bipo_bulk_promptDelay}
        \end{figure}

        In the skin LXe region, where charge signal is absent, the light signal alone can be exploited to identify $\alpha$~particle interactions. The distribution of total collected photons from \ce{^214Po} $\alpha$~decays in nEXO's skin LXe is shown in Fig.~\ref{fig:bipo_bulk_promptDelay}. Most $\alpha$~decays result in a total light detection that is more than 10 times larger than that of a \0 event. The exceptions to this include $\alpha$~decays in the regions with low light collection efficiency above the anode and below the cathode, and $\alpha$s that lose some or all of their energy inside a structural component (e.g. in the copper of the field cage rings). 
    
        If the \ce{^214Po} $\alpha$~decay can be tagged in the skin LXe, then a veto can be applied to reject potential \ce{^214Bi} time-correlated backgrounds.  With an $\alpha$ selection cut on the total scintillation light of $>2\times10^4$ photons, the $\alpha$ tagging efficiency is 98\% when the $\alpha$ starts in the barrel of the skin LXe. Alpha particles that lose part or all of their energy in the detector structure are responsible for the 2\% lost efficiency. Above the anode and below the cathode the tagging efficiency drops to $<$1\% due to the poor light collection efficiency. 49\% of the $\alpha$s that decay from the field ring surface are successfully tagged. Half of them are contained inside the ring material.  %

    \subsection{\label{sec:timeCoins}Analysis of coincident TPC LXe and skin LXe events}
        MS-skin events have a single energy deposit in the TPC LXe with one or more coincident energy depositions in the skin LXe.
        
        These events are typically due to Compton scattering or emission of multiple coincident $\gamma$~rays from a single radioactive decay. If the separate skin interaction is not tagged, these events will contribute to the background. 
   
\begin{figure}[tbp]
            \centering
            \includegraphics[width=0.95\columnwidth]{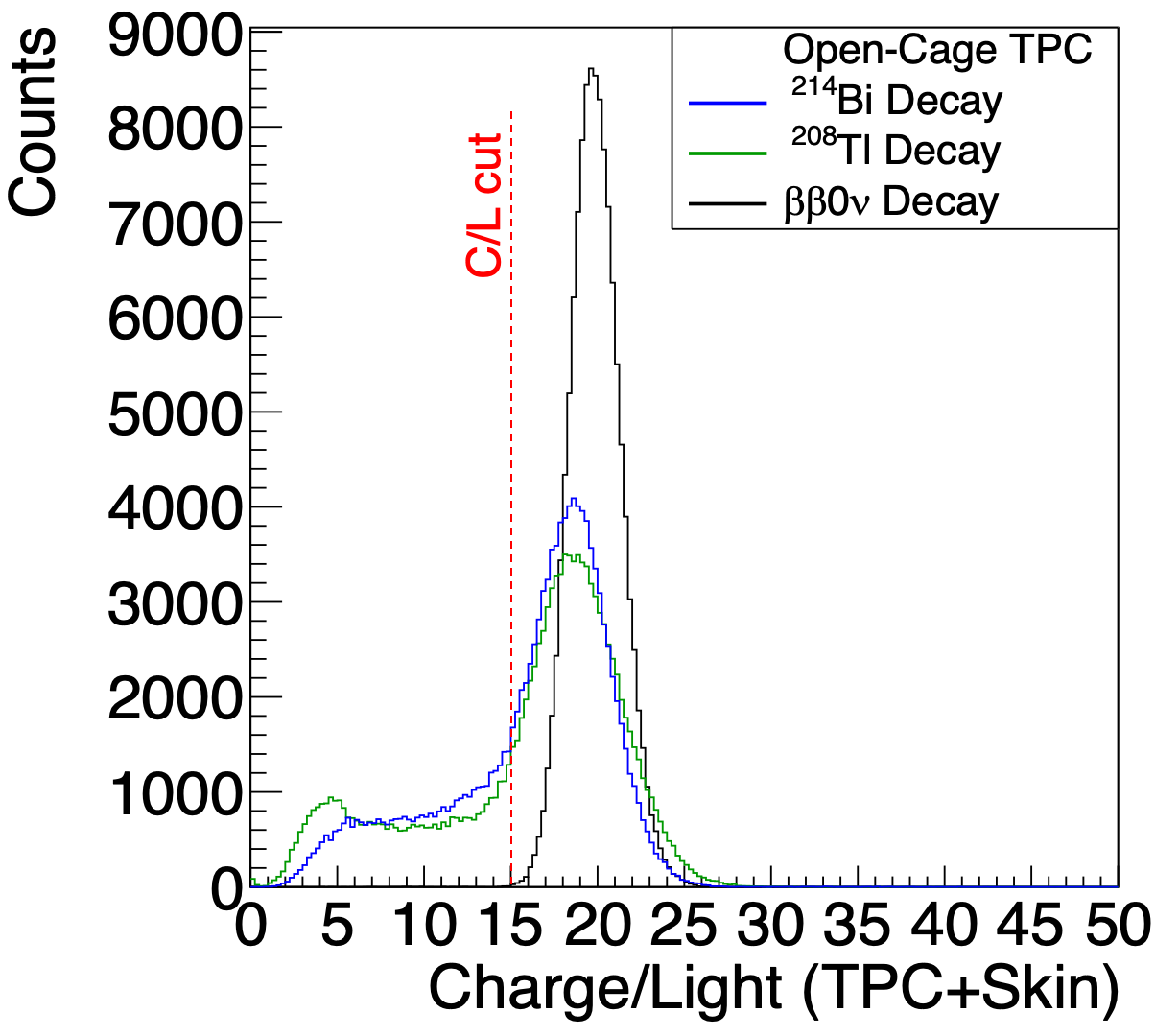}
            \caption{Normalized C/L distribution for SS events in the LXe inner 2000~kg from decays originating from \ce{^214Bi} (blue) and \ce{^208Tl} (green) contamination in the bulk of the copper of the TPC vessel. The distribution for \0 events is also shown (black). 
            The C/L cut in this analysis is shown for reference.}           
           \label{fig:oneDcl}
        \end{figure}

        Coincident energy deposits in the skin LXe and TPC LXe are not resolved by the photosensor timing. The single scintillation signal and single charge cluster resembles a true SS event such as a \0. However, MS-skin events typically have a lower ratio of the charge signal amplitude to the light signal amplitude (``C/L'' ratio) than true SS events. This effect is visible in Fig.~\ref{fig:oneDcl} which shows the distribution of C/L ratios for simulated decays originating in the copper TPC vessel. 
      In the absence of skin interaction, as in the case of \0 events, an almost-Gaussian distribution of C/L values is observed, the width of which is determined by the recombination fluctuations between charge and light production, the individual resolution for each channel, and NEST's parametrization of the energy-dependence of C/L. In contrast, for $\gamma$-ray events which can include skin interactions, a broader distribution of C/L values is seen. A considerable fraction of events, 56\% of \ce{^214Bi} and 69\% of \ce{^208Tl}, were MS-skin events and so have reduced C/L values. For clarity, only the distributions of decays from \ce{^214Bi} (\ce{^238U} chain) and \ce{^208Tl} (\ce{^232Th} chain) have been shown since these are the only  contributors, 48\% and 4\% respectively, to the \0 background near \Q. %
      
\begin{figure*}[tbp]
            \centering
             \includegraphics[width=0.49\textwidth]{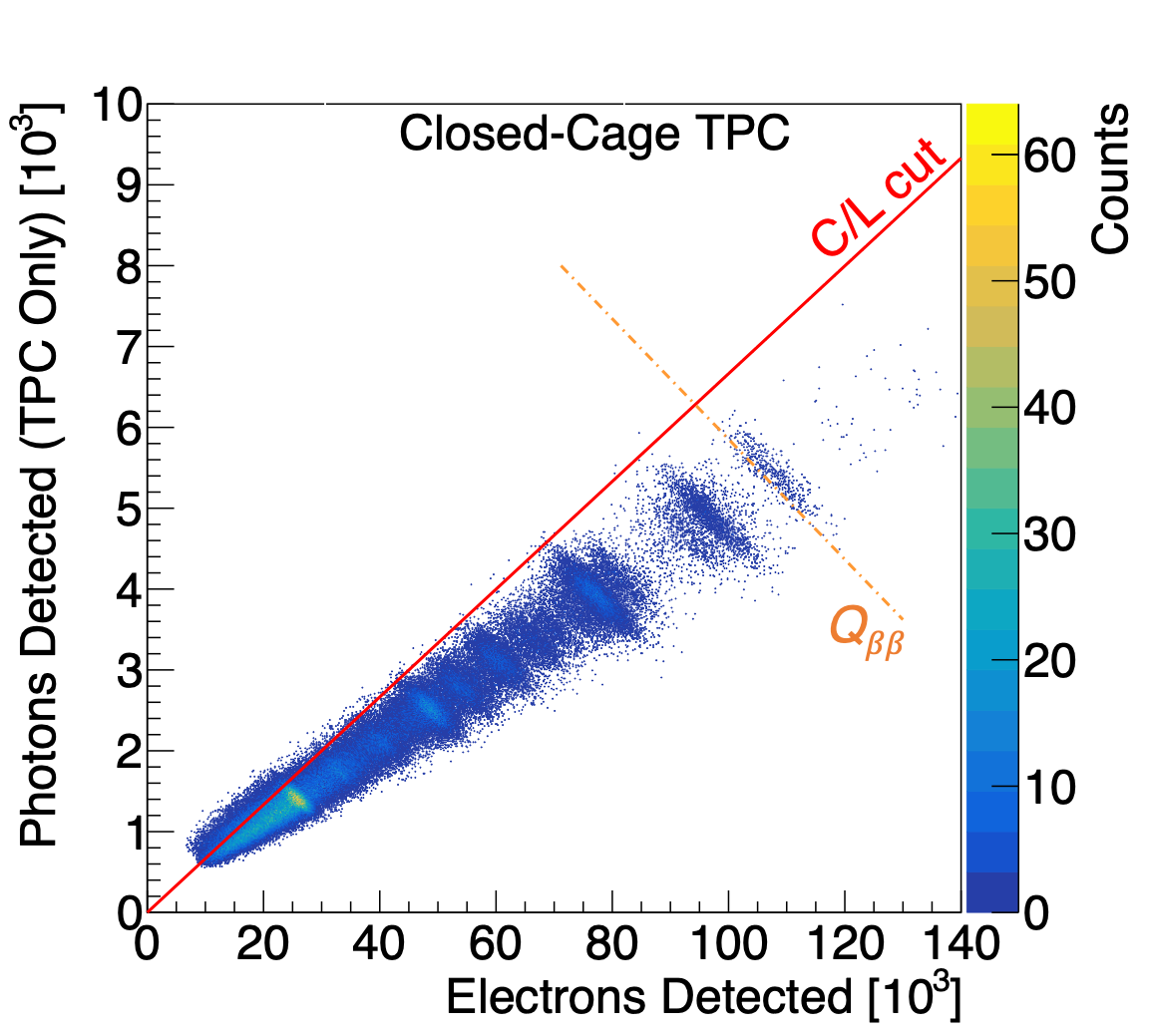}
            \includegraphics[width=0.49\textwidth]{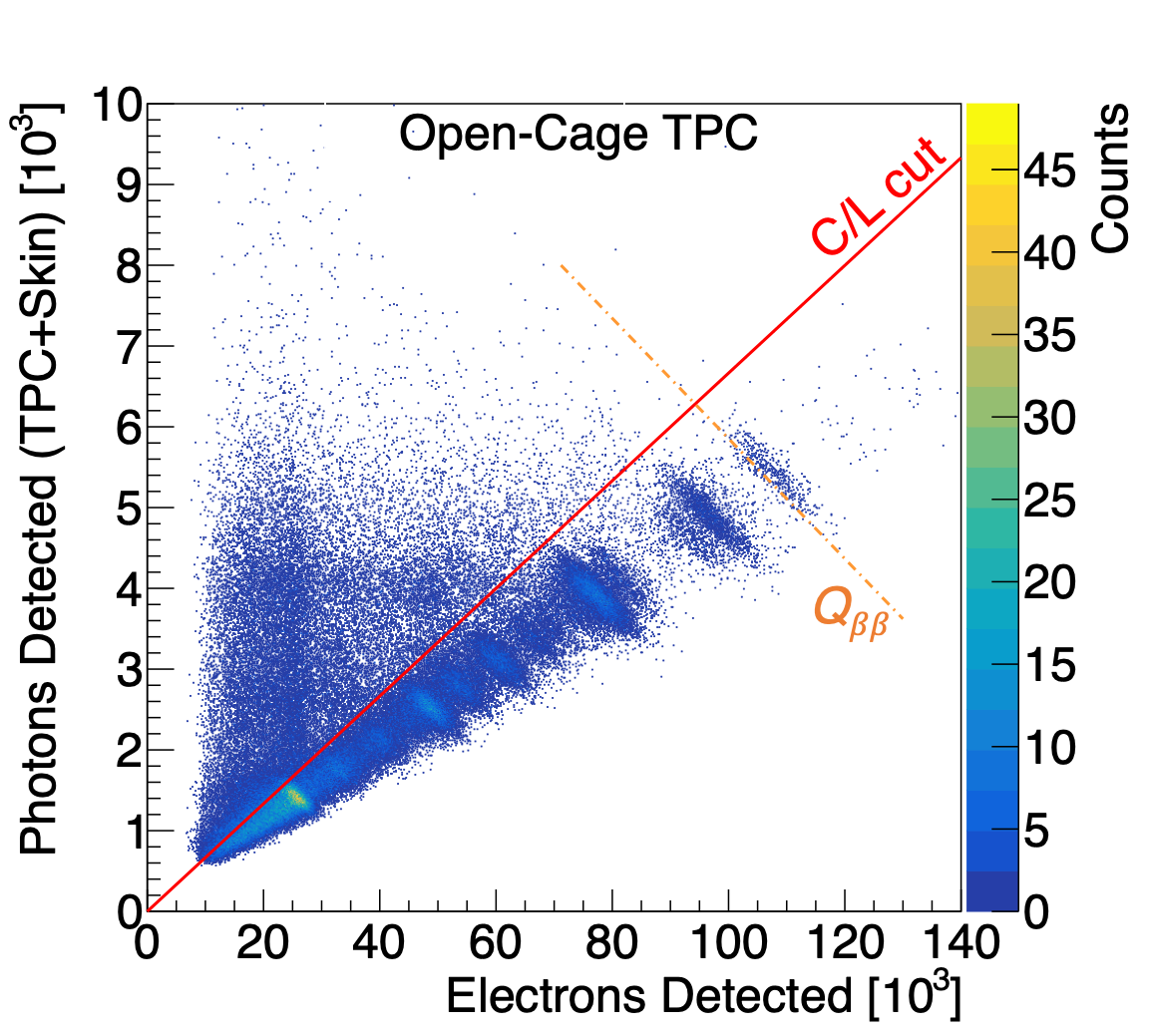}
            \caption{Detected Photons vs Detected Electrons for \ce{^214Bi} decays inside the copper of the TPC vessel. Simulations assumed closed-cage TPC (left) and open-cage TPC (right) in the inner 2000~kg LXe. The \Q~rotated energy value and C/L cut are shown for reference. }
            \label{fig:CvsL-214Bi}
        \end{figure*}

\begin{figure*}[tbp]
            \centering
            \includegraphics[width=0.49\textwidth]{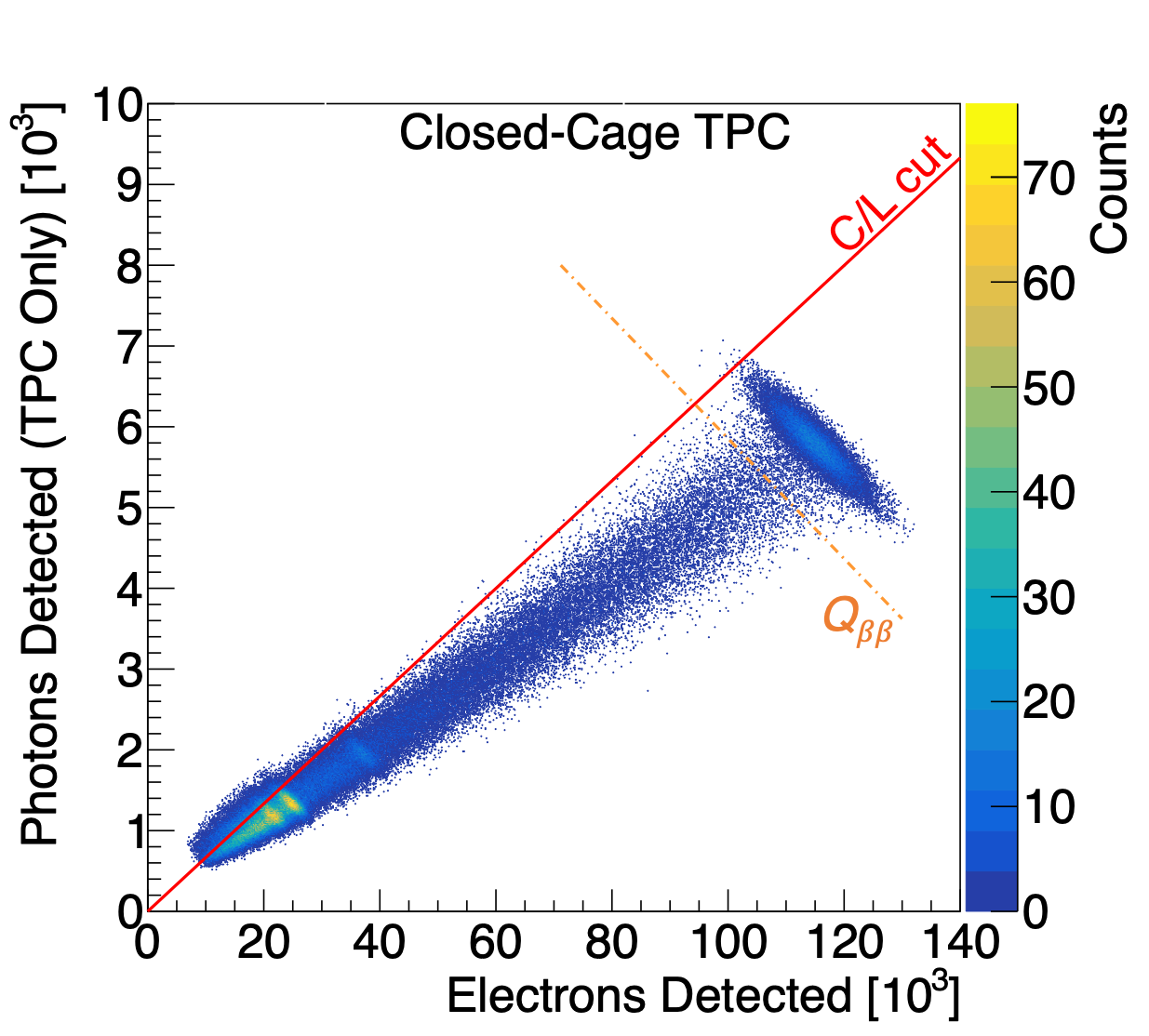}
            \includegraphics[width=0.49\textwidth]{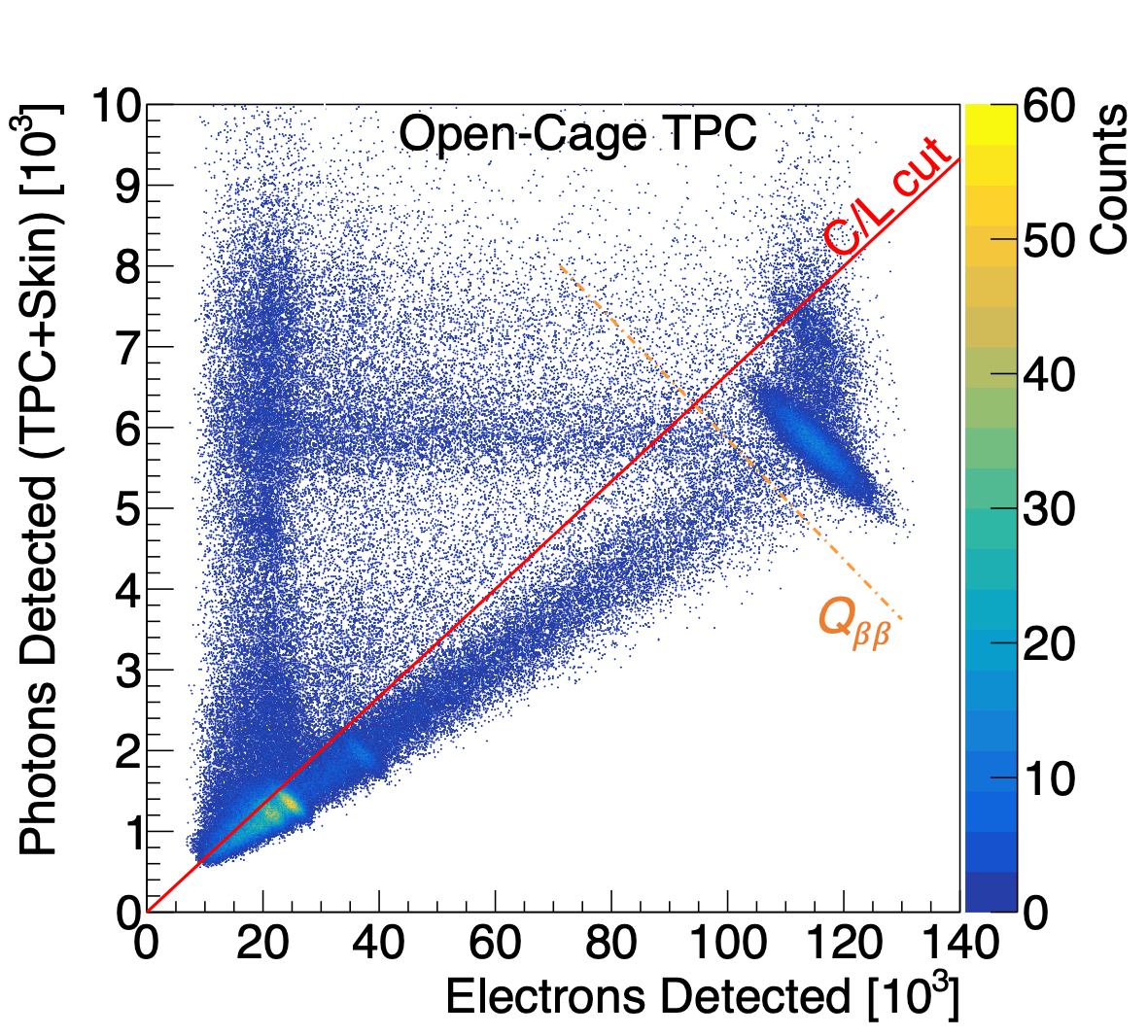}
            \caption{Detected Photons vs Detected Electrons for \ce{^208Tl} decays inside the copper of the TPC vessel. Simulations assumed closed-cage TPC (left) and open-cage TPC (right) in the inner 2000~kg LXe. The value for \Q~in the rotated energy axis and the C/L cut are shown for reference. }
            \label{fig:CvsL-208Tl}
        \end{figure*}

      The effect of the additional light from coincident interactions in the skin can be better appreciated in the scatter plot of photons vs electrons detected, as shown in Fig.~\ref{fig:CvsL-214Bi} and Fig.~\ref{fig:CvsL-208Tl} for simulated \ce{^214Bi} and \ce{^208Tl} decays inside the copper of the TPC vessels respectively. Only events that have a single reconstructed charge cluster, within the inner 2000~kg LXe, are shown.
        In closed-cage TPC (left plots), which was assumed in the previous analyses \cite{Albert:2017hjq,Kharusi:2018eqi}, all SS events lie along the same C/L line, with some variation around that line due to anticorrelated recombination fluctuations. In an open-cage TPC (right plots), many events occur with disproportionately more photons. These extra photons are produced by interactions in the skin LXe.
        
        The features in the distribution of events outside of the anticorrelation band can be understood by considering the two processes responsible for generating events with a single deposit in the TPC LXe in coincidence with one or more deposits in the skin LXe: Compton scattering of a single $\gamma$~ray and coincident $\gamma$~rays.
        
        When a single $\gamma$~ray Compton scatters in the skin LXe volume followed by a photoelectric interaction in the TPC LXe volume, in an open-cage design, the scintillation signals sum to a value dependent on the energy of the initial $\gamma$ ray and the interaction positions, while only the fraction of charge deposited in the TPC LXe is collected. This gives rise to horizontal band structures, like the one around 6000 detected photons in the right plot of Fig.~\ref{fig:CvsL-208Tl}, which corresponds to the 2615 keV $\gamma$~ray from \ce{^208Tl} decay. %
        The resolution of the photon channel is degraded due to the variations of the photon yield and limited collection efficiency in the different skin regions.
        
        Radioactive decays that emit multiple coincident $\gamma$-ray emissions are responsible for events like those in the structure protruding above the spot at $\sim$110,000 detected electrons in the right plot of Fig.~\ref{fig:CvsL-208Tl}. These events arise from the 2615 keV $\gamma$~ray undergoing photoelectric interaction or absorption  in the TPC LXe, with a coincident $\gamma$~ray emitted in the same decay (85\% of the time the 583 keV $\gamma$~ray ~\cite{NDS:2007}) interacting in the skin LXe. This type of events represents 13\% of \ce{^208Tl} decays in the \Q$\pm$FWHM/2 energy window. This feature is not present near the 2448~keV $\gamma$~ray from the \ce{^214Bi} decay because these decays nearly always have a $\gamma$-ray multiplicity of 1. 
      
        As extreme values of the C/L ratio are produced only by backgrounds, events with these values should be rejected. Here, we assess the impact of a simple cut that removes events with C/L $<15$, as shown by the red line in Fig.~\ref{fig:CvsL-208Tl} and Fig.~\ref{fig:CvsL-214Bi}. The value for this C/L cut was defined by requiring $>99\%$ \0~signal acceptance rate. Experimentally, this value would be determined from calibration and \2 data as was done in the recent EXO-200 analysis \cite{PhysRevLett.123.161802}.
        
\begin{figure*}[tbp]
            \centering
            \includegraphics[width=0.49\textwidth]{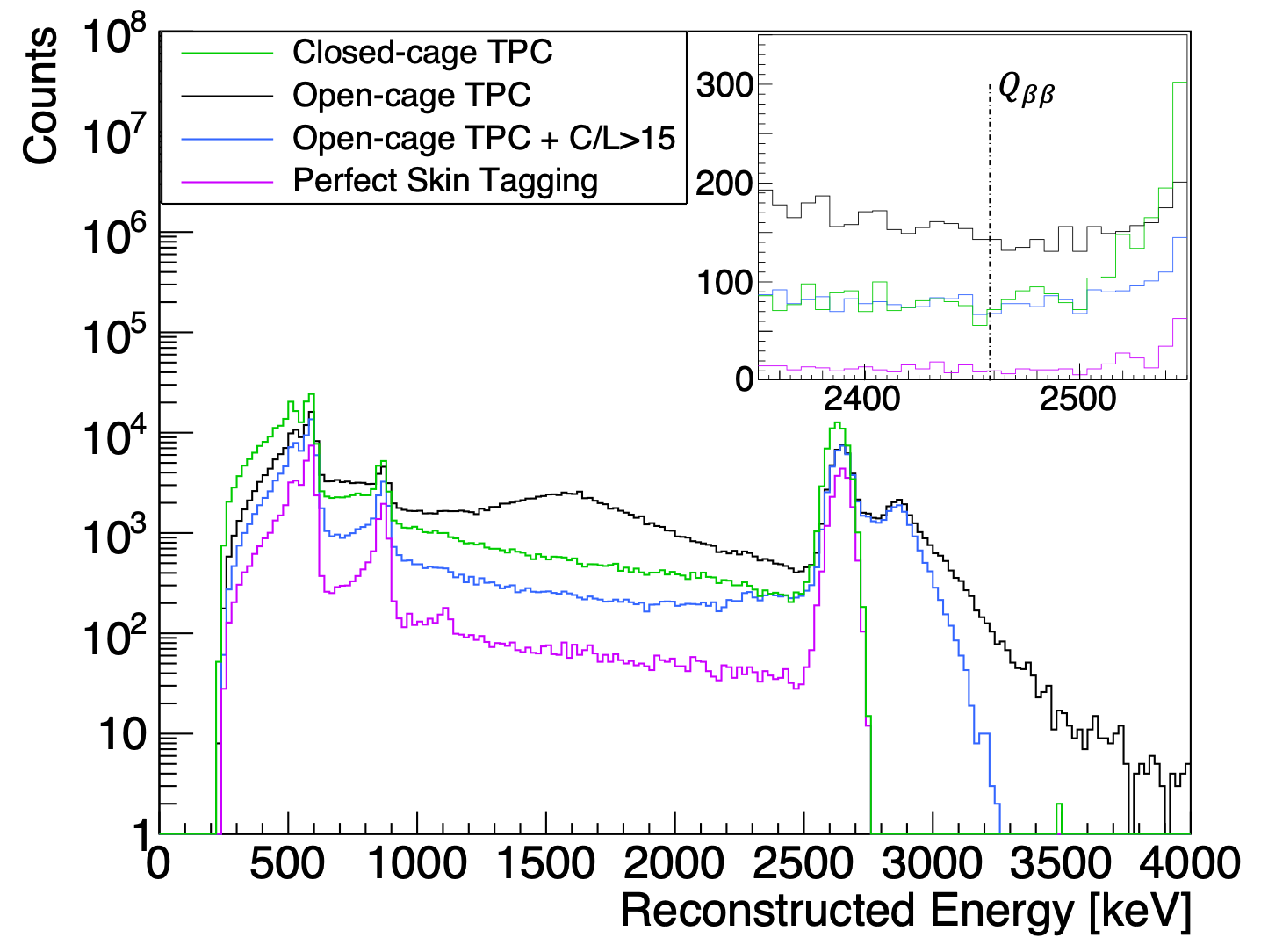}
            \includegraphics[width=0.49\textwidth]{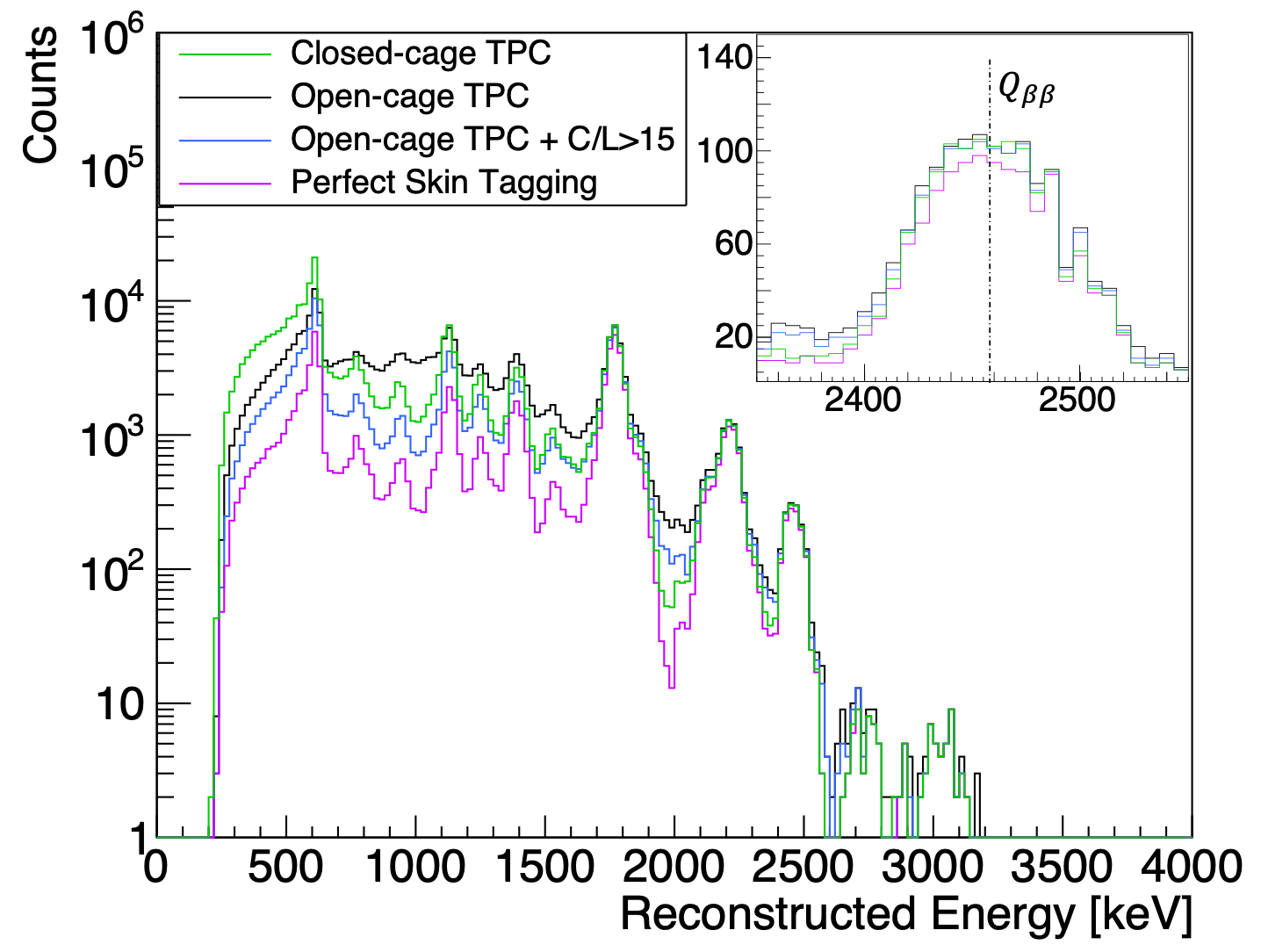}
            \caption{Reconstructed energy of SS events in the inner 2000~kg TPC LXe for \ce{^208Tl} (left) and \ce{^214Bi} (right) decays inside the copper of the TPC vessel, under different conditions.  The distribution obtained for a closed-cage TPC (green), is compared to that obtained in the open-cage TPC with and without removing events with C/L$>15$ (black and blue lines respectively). The results from the ideal case where interactions in the skin LXe could be perfectly reconstructed and identified are also shown (magenta). The insets show a zoom of the distributions near the \ce{^136Xe} \Q.}
            \label{fig:recon-energy}
        \end{figure*}

        The reconstructed energy (using the optimized linear combination of light and charge) is shown in Fig.~\ref{fig:recon-energy} for simulated \ce{^208Tl} and \ce{^214Bi} decays originating in the TPC vessels. Four spectra are overlaid in these plots. Each spectrum shows events with a single reconstructed charge cluster in the inner 2000~kg TPC LXe. The first spectrum shows the energy that would be obtained in a closed-cage TPC. The second spectrum shows the energy observed by an open-cage TPC. In this spectrum, due to coincident skin light the energy can be misreconstructed, giving rise to features like the hump on the right of the \ce{^208Tl} peak. The third spectrum shows the open-cage TPC results after applying the C/L~$> 15$ requirement. Finally, the fourth spectrum applies a cut using MC truth, excluding all events with a skin component. This shows the hypothetical ideal selection which removes all skin-interacting backgrounds. In the region of interest around \Q,~ no significant difference is observed between the different analyses for \ce{^214Bi} decays. On the other hand, MS-skin events account for a large fraction of the background originating from \ce{^208Tl}. This is due to shallow Compton scattering in the skin LXe of the 2615~keV $\gamma$-ray, and to the coincident 583~keV $\gamma$-ray depositing energy in the skin LXe. %
        The addition of coincident skin light to the energy reconstruction impacts the background rate across the entire detector, not only in the inner 2000~kg. Therefore, the relative impact on the expected background contribution in the \Q$\pm$FWHM/2 energy window as a function of volume was also inspected.
        
    For \ce{^208Tl}, a non-linear increase in the SS backgrounds near the \Q~was observed when considering interactions in the TPC LXe volume closer to the edges of the field cage. This behavior is understood by looking at a breakdown by event type of these interactions.
    The background in the inner 2000~kg of LXe is made up of mostly (79\%) events where the with primary $\gamma$~ray Compton scatters in the skin LXe, then deposits the rest of its energy in the TPC LXe. The remainder of events are divided between decays with multiple $\gamma$~rays (13\%), where the primary $\gamma$~ray (2615~keV) deposits energy inside the TPC LXe and a second coincidence $\gamma$~ray deposits energy in the skin LXe, and events where the primary $\gamma$~ray Compton scatters in the TPC LXe, followed by photoelectric interaction or absorption in the skin LXe (8\%). %
    The TPC LXe volume outside of the inner 2000~kg has a different profile of event types. It has a larger (49\%) contribution from those events where the primary $\gamma$~ray scatters first in the TPC LXe then the skin LXe. They are predominately events where the energy deposited in the TPC is near the Compton edge for the 2615~keV $\gamma$~ray, located at 2382~keV. By itself, a Compton scatter at this energy is too far away from \Q~to be included in the background count. However, when including additional light from the subsequent interaction in the skin, these events are misreconstructed near the \Q~value. Because the energy deposited in the skin LXe is typically small ($\sim$~100-200~keV), the extra light is not sufficient to bring the C/L value below the cut threshold. Thankfully, these events predominantly appear in the outer volume of the TPC LXe and have a limited impact on the sensitivity. In the simulation performed for this work, no angular correlation was assumed between coincident $\gamma$-rays from decays such as \ce{^208Tl} \cite{Biedenharn:1953}. This effect was evaluated in the context of EXO-200 calibration and found insignificant \cite{MacLellan:2012}.

\section{\label{sec:results}Results and Discussion}
 
    We now turn our attention to the impact that the open-cage design has on the background in nEXO. As discussed earlier, the nEXO \0 sensitivity estimations previously assumed that
    for interactions in the skin, only $\alpha$~particles could be detected. All signals from $\gamma$-ray interactions in the skin were neglected. In doing so, energy misreconstructions were not considered nor were possible background reductions from exploiting charge-to-light features. %
    The background resulting from the prior simplified analysis is summarized in the first column of Table~\ref{tab:results}. This summary includes only the main background components in nEXO relevant to the analysis of interactions in the skin LXe. These include interactions from the \ce{^238U} and \ce{^232Th} decay chains from bulk contamination of the materials in the TPC vessel and internal components. Contributions from daughters of dissolved \ce{^222Rn} are also included. Values are normalized to the total background budget. Backgrounds were estimated near \Q~for the central region of the detector (inner 2000~kg), since this region dominates the detector sensitivity \cite{Albert:2017hjq}.  
        
    The second column of Table~\ref{tab:results} shows the effect on the background budget of exploiting the light collected from interactions in the skin LXe in an open-cage TPC, calculated using the method discussed in the previous section. The $C/L<15$ cut was applied to reject interactions with a skin component while retaining more than 99\% of \0 events. %
    For identification of Bi-Po events in the skin, the analysis considered a veto of one second before any $\alpha$~decay in the skin. The veto time is significantly longer than the \ce{^214Po} half-life of 163~$\mu$s, and so the Bi-Po tagging efficiency is dominated by the ability to observe and tag the $\alpha$~decay in the skin. Even when an $\alpha$ is not visible in the skin, sometimes the light signal from the coincident $\beta$ particle emitted during the \ce{^214Bi} decay is sufficient for rejection via the C/L cut.  The analysis resulted in a Bi-Po rejection efficiency of 76\% and 55\% for events originating in the skin LXe and field ring surfaces, respectively. 
    
    These results confirm that the background estimated under the skin-interaction assumptions in the prior study were slightly conservative when compared to the results from this new analysis that fully exploits the open-cage design of nEXO. The previous assumption that coincident $\gamma$-ray interactions in the skin would not impact the total backgrounds proved largely correct in this more detailed analysis. This minimal impact stems from the small fraction of \ce{^214Bi} decays that have multi-sited interactions involving the skin and the efficacy of the C/L ratio cut at removing \ce{^208Tl} skin-interacting backgrounds. This study's detailed accounting of skin Bi-Po events shows the background from \ce{^222Rn} dissolved in LXe can be rejected more efficiently than assumed in the simplified analysis.   
 
    Finally, we consider the ideal case in which the detector can identify with 100\% efficiency all MS-skin events and \ce{^214Po} $\alpha$ emission in the skin. This is shown in the last column of Table~\ref{tab:results}. In this case, the \ce{^232Th} component is more than halved, while the \ce{^238U} component is largely unchanged. The decrease in the \ce{^232Th} component is because of shallow Compton scatter in the skin of the 2615~keV $\gamma$~ray and decays with $\gamma$-ray multiplicity larger than one. These are MS-skin whose skin interactions are too low in energy to be tagged under the assumptions in the non-ideal, open-field-cage analysis, but which are caught in this ideal scenario.
    The additional improvement in the \ce{^222Rn} background is primarily due to rejection of Bi-Po decays in the LXe regions (above the anode, below the cathode) where the light collection efficiency is small ($\sim$0.1\%) and therefore not taggable in the non-ideal analysis. 

    \begin{table*}[tbp]
        \centering
        \renewcommand{\arraystretch}{0.98}
        \begin{tabular}{lccc}
        \hline
        \hline
            & \multicolumn{3}{c}{\textbf{Background Fraction}} \\ 
        \hline
            \textbf{}          & \textbf{Simplified}  & \textbf{Detailed Open-cage}     & \textbf{Perfect Skin} \\ 
            \textbf{}          & \textbf{Analysis~\cite{Albert:2017hjq}}      & \textbf{TPC Analysis}       & \textbf{LXe Tagging} \\ 
        \hline
            \ce{^238U} TPC Vessel and Internals   & $0.48$   & $0.45$   & $0.43$\\
            \ce{^232Th} TPC Vessel and Internals  & $0.039$  & $0.039$  & $0.015$\\
            \ce{^222Rn} Skin LXe            & \multirow{2}{*}{$0.043$} & $0.0025$ & $0.0015$\\
            \ce{^222Rn} Field Ring Surfaces &    & $0.026$  & $0.019$\\
            All Other Backgrounds           & $0.44$ & $0.43$  &  $0.42$ \\
   
        \hline
            Total Backgrounds & 1.00 & 0.95 & 0.88 \\
        \hline
        \hline
        \end{tabular}
        \caption{Background contributions from SS events to the \Q$\pm$FWHM/2 in the inner 2000~kg LXe. Only the main backgrounds relevant to the skin LXe are shown. All values are normalized to the total background counts at 90\%CL for all regions in the simplified closed-cage analysis (first column). The values in the second column are for the detailed analysis that considered the effects of an open-cage design. Results in the last column are obtained under the ideal case where interactions in the skin LXe could be perfectly reconstructed and identified.}           
        \label{tab:results}
    \end{table*}
 
\section{\label{sec:conclusion} Conclusions}
 
This study has established that an optically-open field cage design introduces some complexity in the event reconstruction of a liquid xenon TPC but can be used to identify and reject skin-interacting backgrounds. Time-correlated Bi-Po backgrounds can be removed by identifying light signals from the \ce{^214Po} decays in the skin and on nearby surfaces. Coincident backgrounds are not significantly different between open-cage and close-cage designs because \ce{^214Bi} backgrounds rarely interact in the skin and \ce{^208Tl} skin-interacting backgrounds can be rejected using a simple cut. Together, these result in a 5\% reduction in background over the estimate in \cite{Albert:2017hjq}. This reduction does not lead to an appreciable improvement in nEXO's sensitivity to \0 \ce{^136Xe} half-life, as nEXO's half-life sensitivity ($T^{0\nu}_{1/2}$) only scales with the background rate in the inner 2000~kg as ($B$) as $T^{0\nu}_{1/2}\propto{B^{-0.35}}$. 

Continued improvements in the ability to reconstruct and tag the light in the skin LXe could provide additional background reduction, up to 12\% in the ideal case of perfect skin tagging. Better understanding of event identification strengthens nEXO's ability to assess and control its backgrounds. Furthermore, the ability to better identify skin LXe interactions may become important during calibration measurements due to pile-up. nEXO plans to calibrate the detector spatial light-collection efficiency using a set of intense $\gamma$-ray sources placed outside of the TPC vessel. In order to reach a sufficient count rate of ``deep'' events in the inner region of the LXe volume, an interaction rate of up to 1.6~kHz is expected in the detector. The majority of those interactions happen in the outer regions of xenon, including the skin LXe. In this situation, rejection of interaction with a skin component can help in isolating the deep events necessary for the calibration.  One approach that can be tried consists of exploiting the spatial distribution of collected photons on the SiPMs. The highly concentrated pattern from a skin LXe $\gamma$~ray may provide a recognizable feature when compared to the diffuse pattern from a centralized deep event.

\section*{Acknowledgements}
Support for nEXO comes from the the Office of Nuclear Physics of the Department of Energy and NSF in the United States, from NSERC, CFI, FRQNT, NRC, and the McDonald Institute (CFREF) in Canada, from IBS in Korea, from RFBR (18-02-00550) in Russia, and from CAS and NSFC in China. LLNL-JRNL-814563

\bibliography{ms-skin}

\end{document}